\documentclass{osa-article}

\journal{oe}
\usepackage{siunitx}
\usepackage{graphicx}
\usepackage{subcaption}
\articletype{Research Article}
\begin{document}

\title{Nanofiber-based high-Q microresonator for cryogenic applications}

\author{Johanna H\"utner,\authormark{1} Thomas Hoinkes,\authormark{1},Martin Becker,\authormark{2}, Manfred Rothhardt,\authormark{2}, Arno Rauschenbeutel,\authormark{1,3,$\dagger$} and Sarah M. Skoff\authormark{1,*}}

\address{\authormark{1}Vienna Center for Quantum Science and Technology, Atominstitut, Technische Universität Wien,
Stadionallee 2, 1020 Vienna, Austria\\
\authormark{2}Institute of Photonic Technologies IPHT, Albert-Einstein-Strasse 9, 07745 Jena, Germany\\
\authormark{3}Department of Physics, Humboldt-Universität zu Berlin, 10099 Berlin, Germany}

\email{\authormark{*}sarah.skoff@tuwien.ac.at,\authormark{$\dagger$}arno.rauschenbeutel@hu-berlin.de} %% email address is required



\begin{abstract}
We demonstrate a cryo-compatible, fully fiber-integrated, alignment-free optical microresonator. The compatibility with low temperatures expands its possible applications to the wide field of solid-state quantum optics, where a cryogenic environment is often a requirement. 
At a temperature of 4.6 K we obtain a quality factor of $\mathbf{(9.9 \pm 0.7) \times 10^6}$. In conjunction with the small mode volume provided by the nanofiber, this cavity can be either used in the coherent dynamics or the fast cavity regime, where it can  provide a Purcell factor of up to 15. Our resonator is therefore suitable for significantly enhancing the coupling between light and a large variety of different quantum emitters and due to its proven performance over a wide temperature range, also lends itself for the implementation of quantum hybrid systems.
\end{abstract}

%%%%%%%%%%%%%%%%%%%%%%%%%%  body  %%%%%%%%%%%%%%%%%%%%%%%%%%
\section{Introduction}
Achieving efficient coupling between quantum emitters and light is an important goal of research in quantum communication and computation, sensor applications and fundamental studies. Such an efficient interaction can for example be realized by means of an optical cavity or by reducing the mode area of the light field to the size of the emitter's interaction cross-section. Optical nanofibers offer such a strong transverse confinement of the light field.This makes them a versatile tool for interfacing different types of emitters such as atoms 
\cite{vetsch_optical_2010,goban_demonstration_2012,hafezi_atomic_2012}, single molecules \cite{skoff_singlemolecules_2018}, quantum dots \cite{yalla_efficient_2012,fujiwara_highly_2011,shafi_quantumdot_2018} and color centers in diamond \cite{liebermeister_tapered_2014,fujiwara_ultrathin_2015}. For many proof-of-principle experiments, cold atoms were the emitters of choice as they represent a well-controlled and isolated system. However, the experimental overhead for a cold-atom set-up is significant and solid-state emitters are much more suitable for practical and scalable platforms for quantum networks or nanosensors \cite{aharonovich_review_2016,awschalom_review_2018}. 

Yet, the optical transitions of solid state emitters also couple to the phononic degrees of freedom of the system, leading to dephasing and inelastic scattering. In order to avoid this problem, the phonons of the system have to be frozen out and the branching ratio of emission into the coherent zero-phonon line has to be maximized. Further, cryogenic temperatures may be required to be able to spectrally address individual solid state emitters in the case, where many are present in the same host system \cite{skoff_singlemolecules_2018}. This calls for a cryo-compatible optical microresonator with high quality factor, Q, that selectively accelerates the desired optical transition via the Purcell effect \cite{benedikter_cavity-enhanced_2017,fujiwara_coupling_2012,wang_cavity_2019, kaniber_efficient_2007}. Here, we show that a fully fiber-based optical microresonator that consists of a tapered optical fiber with two integrated fiber Bragg gratings (FBGs)\cite{Othonos,guo_cryogenic_2012}, as demonstrated in \cite{wuttke_nanofiber_2012}, can be employed at cryogenic temperatures. In particular, we confirm that a high Q factor prevails after contact gas-cooling from room temperature to liquid helium temperature. Consequently, the resonator is still compatible with reaching the strong coupling regime. Furthermore, for usage at cryogenic temperatures, the alignment-free character of our resonator represents an advantage over modular resonators which tend to misalign when being cooled down.

\section{Experimental methods}	
To obtain the in-fiber resonator, two FBGs with similar transmission spectra are laser written into a commercial single mode fiber SM800 \cite{lindner_thermal_2009}. Along the fiber, the gratings are separated by \SI{20}{mm}. In this way, a Fabry-Perot-like resonator for a center wavelength of \SI{852}{nm} is created \cite{wuttke_nanofiber_2012,kato_cavity_2015}.
To achieve a small mode volume of the resonator and to enable evanescent coupling of quantum emitters to this cavity mode, the fiber section between the two FBGs is tapered using a homebuilt fiber pulling rig \cite{warken_fiber_2008}. The tapered optical fiber features a nominal waist length of \SI{3}{mm} and diameter of \SI{500}{nm}. In the taper section, the weakly guided LP$_{01}$ mode of the standard single mode optical fiber is adiabatically transformed into the strongly guided HE$_{11}$ mode of the nanofiber waist and back and the small waist diameter guarantees single mode operation for wavelengths $\lambda>691$ nm. After the tapering process, the gratings are separated by \SI{93}{mm}. 
In view of the guided resonator modes, this type of cavity is completely alignment-free.
\begin{figure}[ht]
	\centering
	\includegraphics[width=0.6\linewidth]{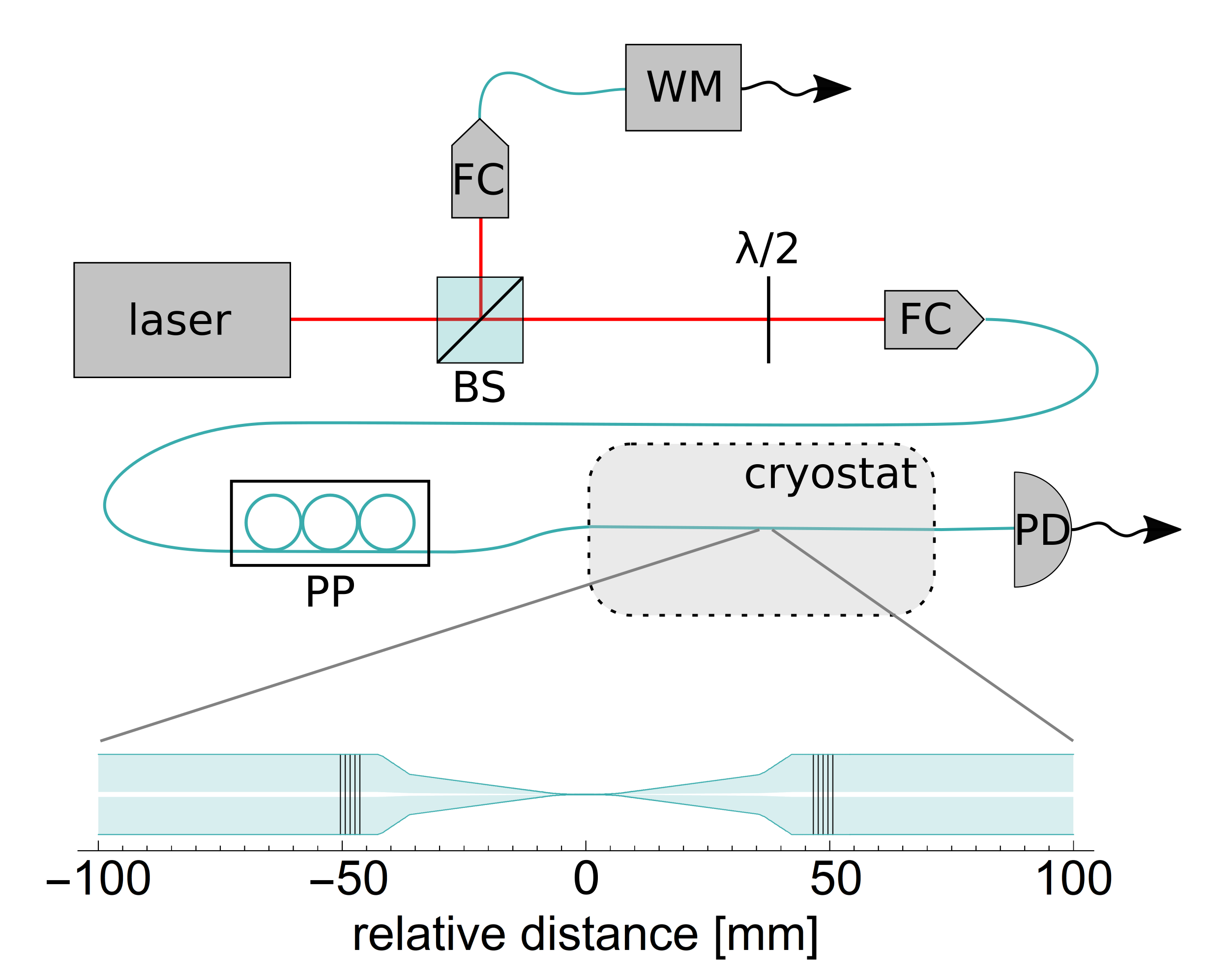}
    
    \caption{Optical setup with a closeup of the nanofiber-based resonator; BS - beam splitter, FC - fiber coupler, WM - wavemeter, PP - polarization paddels, PD - photodiode} 
	\label{fig:optical_setup}
\end{figure}
To cool the cavity, it is mounted in a copper probe chamber inside a liquid helium bath cryostat. For characterization of the resonator prior to tapering, it is clamped onto a teflon mount. After tapering, the fiber resonator is fixed using Stycast epoxy (Loctite Stycast 2850FT with CAT23LV) onto a silica fiber mount on one side and a piezo element that sits on a copper support on the other side. This method of mounting the resonator in principle enables strain-tuning of the cavity
%\cite{james2002}
using the piezo.This is however often not a necessary requirement for coupling solid-state emitters to the resonance of a cavity as the emitters often exhibit a large inhomogeneous broadening due to their nanoenvironment and thus a suitable emitter can be picked out from the ensemble \cite{Rattenbacher2019,gregor_microresonator_2009}. First measurements did not show any effect of the piezo, most probably due to insufficient pre-straining of the tapered fiber. If the fiber is pre-strained enough \cite{holleis_strain_2014} before cool-down, the maximum displacement of the piezo of about $45 \mu$m at liquid helium temperatures is enough to scan the cavity over many free spectral ranges. Alternatively, the piezo can also be mounted on the silica fiber holder directly to avoid any differential length contraction. The fiber mount itself is attached to a copper support with chamfered edges to prevent scratching of the bare optical fiber that could then cause it to break. To achieve efficient cooling of the entire fiber cavity, the probe chamber is slowly evacuated and backfilled with about \SI{50}{mbar} of helium buffer gas before cool-down. The cryostat is then filled with liquid nitrogen and liquid helium and everything is left to thermalize. The temperature of the sample chamber wall is continuously monitored by a calibrated ruthenium-oxide temperature sensor. 

 To optically characterize the cavity, a tunable laser (Velocity TLB-6716) is launched into the optical fiber that is led into the cryostat through a teflon feedthrough \cite{teflon_feedthrough_1998} and spliced to the fiber-based cavity. The transmission of the cavity is measured using a photodiode. Part of the laser beam that is coupled into the optical fiber is split off and coupled into a wavemeter (HighFinesse), which continuously monitors the laser frequency with a relative precision better than $3 \times 10^{-7}$. The optical setup is depicted in Fig. \ref{fig:optical_setup}, where the inset shows a closeup of the nanofiber-based resonator with its nominal radius-profile.

\section{Results and discussion}
\begin{figure}[ht]
	\centering
	\includegraphics[width=0.7\linewidth]{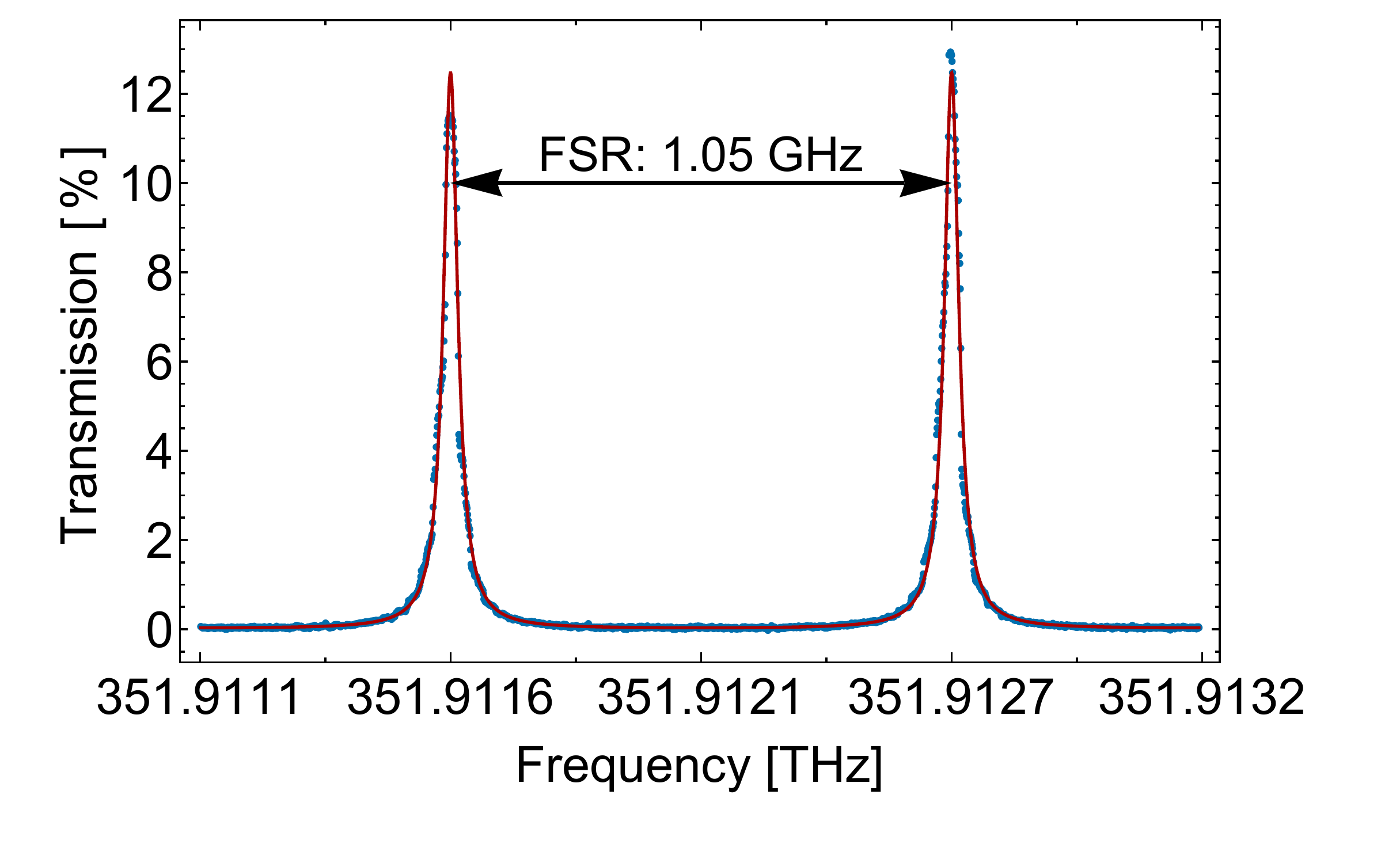}
    
    \caption{Transmission spectrum over one FSR including an Airy function fit. The transmission is normalized to the off-resonant transmission outside the stop band.} 
	\label{fig:fsr}
\end{figure}

\begin{figure}[ht]
	\centering
	\includegraphics[width=0.7\linewidth]{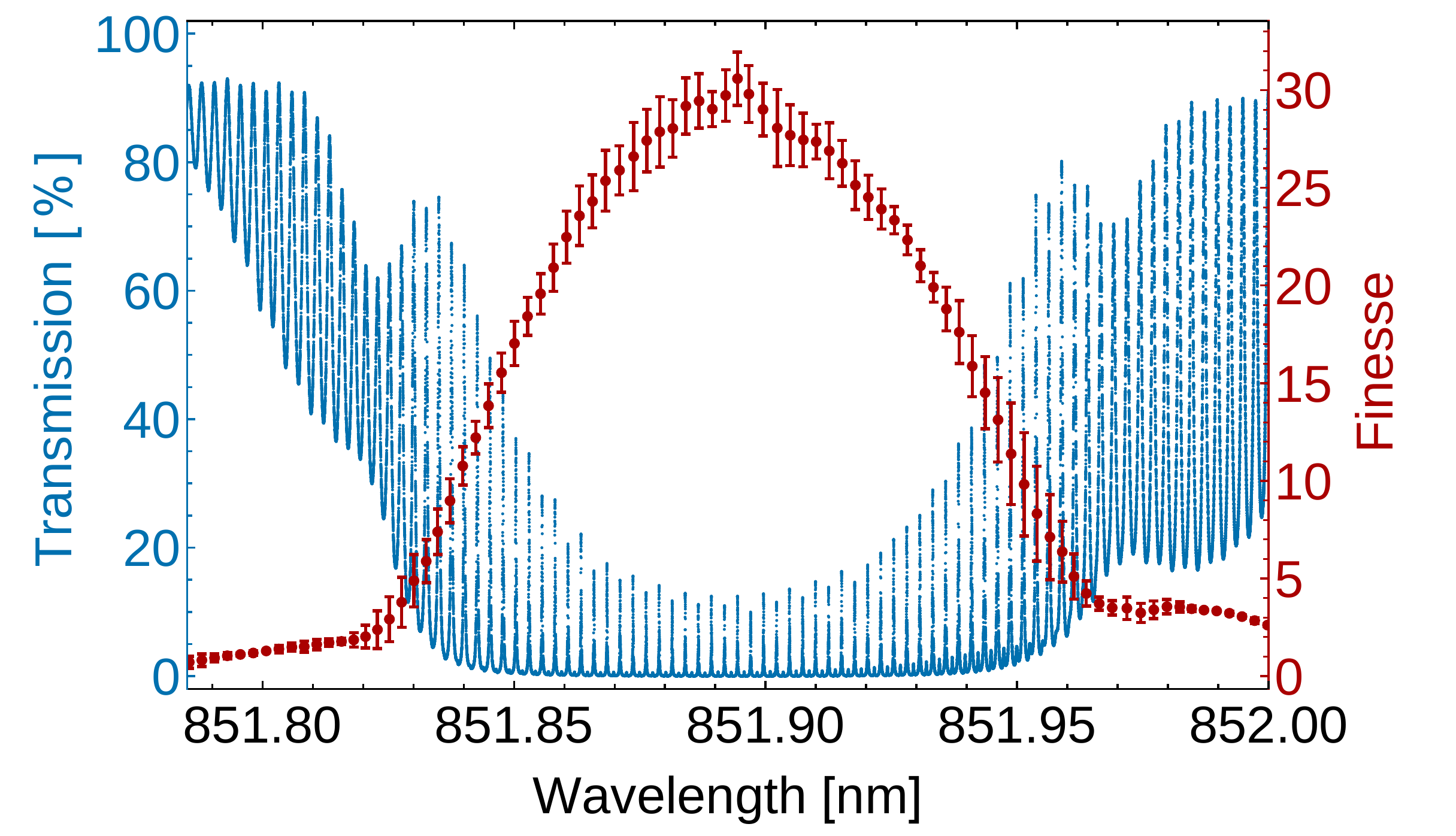}
    \caption{Transmission and finesse of the nanofiber-based resonator as a function of laser detuning at \SI{4.6}{K}. The cavity modes show up as transmission peaks within the \SI{0.2}{nm} wide stop band of the FBGs. The transmission is normalized to the off-resonant transmission outside the stop band.} 
	\label{fig:transmission}
\end{figure}

Figure \ref{fig:fsr} shows the transmission of the cavity covering one free spectral range (FSR) at \SI{4.6}{K}. As the nanofiber-based resonator is inherently birefringent, the input polarization has to be tuned using polarization paddles such that only one resonance per FSR is observed \cite{wuttke_nanofiber_2012}.
The $\lambda/2$-plate (Fig. \ref{fig:optical_setup}) then enables the selective excitation of the two orthogonal quasi-linearly polarized modes of the resonator. By fitting Airy functions to transmission scans at different laser frequencies, the finesse as a function of frequency is obtained. This procedure is repeated several times and the mean value of the finesse is plotted together with the transmission for a wide wavelength range in Fig. \ref{fig:transmission}. We define the center wavelength of the resonator, $\lambda_\text{c}$, as the point of best overlap of the reflection bands of the two FBGs and hence, the point of maximum finesse of the resonator. 

At a temperature of \SI{4.6}{K} we measure a maximum finesse of $29.4 \pm 1.3$ for the nanofiber-based resonator. The spectral position of the corresponding transmission peak corresponds to the center wavelength of the resonator, yielding $\lambda_\text{c} =$ \SI{851.8944}{} $\pm$ \SI{0.0025}{nm} ($\nu_\text{c} =$ 351.9127 $\pm$ \SI{0.0011}{THz}), where the error is one FSR.
When the FBGs are cooled, thermal contraction and temperature-induced refractive index change cause a shift of the center frequency of the resonator. However, for a stable cavity that shall be on resonance with certain quantum emitters, a good knowledge of the wavelength shift at cryogenic temperatures is crucial. We measure a center wavelength of $852.5555 \pm 0.0026$ nm at room temperature, and hence a temperature-dependent wavelength shift of $-0.6611 \pm 0.0051$ nm upon cooldown. The room temperature measurements of the center wavelength before cooling down and after warming up the resonator again are shifted by 0.0209 nm. Before cooldown we measure a lower maximum finesse of $16.0 \pm 0.4$ compared to after warming up of the resonator, where a maximum finesse of $31.6 \pm 2.5$ is measured. The lower finesse value before cooldown may stem from torsion or strain that shifts the reflection band of one of the FBGs a little, leading to a slight shift of the center wavelength and, due to the altered overlap of the two reflection bands, to a lower finesse. The strain associated with the observed wavelenght shift can be calculated \cite{Othonos} to be $\epsilon = 3.1 \times 10^{-5}$. This effect is diminished upon cooldown and we obtain an equally high finesse value at 4.6 K and at room temperature after warming up of the resonator. This means that the shift obtained from these two measurements is solely due to the temperature change and hence this measurement of the temperature-dependent wavelength shift is more reliable for investigating the thermo-optic effect in silica. More details on the temperature-dependent wavelength shift and the thermo-optic coefficient of the FBG resonator can be found in the appendix.

Generally, the tapering process itself can reduce the finesse of the infiber-cavity due to additional taper transmission losses and due to possible thermally induced decrease of the FBGs reflectivities. In our case we measure an initial finesse value of $177.7 \pm 26.4$ at 4.6 K before tapering the fiber section between the two FBGs. The finesse of the resonator can be written  in terms of the FBG reflectivities $R_1$ and $R_2$ and a single pass cavity transmission $T_c$ as   
\begin{equation}\label{eq:finesse}
\mathcal{F} = \frac{\pi \sqrt[4]{R_1 R_2} \sqrt{T_c}}{1-T_c\sqrt{R_1 R_2}}. 
\end{equation}
Assuming that the propagation losses of the unprocessed fiber are negligible ($T_c=1$) and that both mirrors have the same reflectivity, we obtain $R=98\% $ from the measured finesse value. The transmission of a Fabry-Perot cavity with a phase difference  $\delta$ between consecutive transmitted beams is given by
\begin{equation}\label{eq:transmission}
T = \frac{(1-R_1)(1-R_2)T_c}{(1-T_c \sqrt{R_1 R_2})^2} \frac{1}{1+(2 \mathcal{F}/\pi)^2\sin^2{\delta/2}}.
\end{equation}
For the nanofiber-based resonator at 4.6 K, we measure a finesse of 29.4 and a maximum transmission on resonance of $T/T_c = 0.14$. Solving equations \ref{eq:finesse} and \ref{eq:transmission} simultaneously yields a reflectivity per mirror of $R_1 = R_2=96\%$ and a nanofiber transmission of $T_c = 93\%$. This is lower than the final nanofiber transmission of $98\%$ monitored during the heat and pull process and may be due to dust that accumulated on the fiber waist. In principle, a higher fiber transmission should be achievable as transmissions of up to 99.95\% have been demonstrated \cite{hoffman_ultrahigh_2014}.% The off-resonance transmission of the FBGs was measured using a white light source and a spectrometer with a resolution better than 1.3 nm to be $>85$\%, in the range from 750 to 850 nm and 854 to 1000 nm. 

At a temperature of \SI{4.6}{K}, the maximum finesse of 29.4 in conjunction with an FSR of 1.05 $\pm$ \SI{0.06}{GHz}, yields a Q-factor of our nanofiber-based resonator of $(9.9 \pm 0.7) \times 10^6$. A figure of merit for the performance of a resonator with respect to the coupling efficiency between the light field and a quantum emitter is the cooperativity $C = g^2/(2 \gamma_0 \kappa)$, where $2 g = \sqrt{2 \mu^2 \omega/( \hbar \epsilon_0 V)}$ is the single photon Rabi frequency with the dipole moment $\mu$, the effective mode volume $V$ and the free space permittivity $\epsilon_0$. Furthermore, $2 \kappa = \omega/Q$ is the cavity decay rate and $ 2 \gamma_0$ the free space spontaneous emission rate of the emitter. $g$ and $\kappa$ scale with the cavity length $L_{\text{c}}$ as $\propto 1/\sqrt{L_{\text{c}}}$ and $\propto 1/L_{\text{c}}$, respectively, while $\gamma_0$ is independent of the cavity length. 

Thus, $C$ is independent of $L_{\text{c}}$ and by choosing the cavity length appropriately, a cavity, given $C>1$, can operate in the coherent dynamics or the Purcell regime \cite{wuttke_nanofiber_2012}. For a two-level emitter, the cooperativity is related to the enhancement of the spontaneous emission by $ 2C\!=\!F_{\text{P}}\!=3 \lambda^3/(4 \pi^2) Q/V$, where $F_{\text{P}}$ is the Purcell factor \cite{photonics_2015}. 
Assuming perfect overlap between the dipole moment of an emitter on the surface of the nanofiber and the quasilinearly-polarised field of the nanofiber gives a minimum effective mode volume of $4.9 \times 10^4 \lambda^3$, where the mode volume is the effective mode area \cite{skoff_singlemolecules_2018,warken_ultra-sensitive_2007} times the length of the cavity determined by the free spectral range. The mode area is defined as $A_{\text{eff,surf}} = P/(I_{\text{surf}}\, (\mathbf{\hat{d}}\cdot \mathbf{\hat{e}_{\text{surf}}})^2)$, where $I_{\text{surf}}$ is the surface intensity, $\mathbf{\hat{e}_{\text{surf}}}$ the polarization vector at the position of the emitter and $\mathbf{\hat{d}}$ the unit vector of the dipole moment of the solid-state emitter.For a perfect overlap between polarization of the light field and the transition dipole moment of the quantum emitter, the maximum cavity Purcell factor is 15 on the surface of the nanofiber. Assuming an orientation-averaged dipole instead, gives an average cavity Purcell factor of 5. In practice, achieving maximum alignment between the dipole moment of a solid-state emitter in a nano-solid and the polarisation vector of the cavity may prove challenging. However, if an amorphous solid is used \cite{faez_coherent_2014}, the dipole moments of individual emitters differ in direction and the most suitable emitter can be chosen. 

When calculating the channeling efficiency into the nanofiber-based cavity in the fast cavity regime, one also has to consider the inherent Purcell enhancement $P_{\scriptscriptstyle \text{TOF}} = \gamma_{\text{total}}/\gamma_0$ induced by the optical nanofiber without mirrors \cite{nayak_nanofiber_2018}. This amounts to 1.57 in our case. Here, the total scattering rate of the emitter $\gamma_{\text{total}} = \gamma_{\text{guided}}+\gamma_{\text{rad}}$, is the sum of the scattering rate to guided modes of the nanofiber without mirrors and to radiation modes. The channeling efficiency for a nanofiber-based cavity is then given by \cite{lekien_cavity_2009}
\begin{equation}
\eta_c = \frac{\eta P_{\scriptscriptstyle \text{TOF}} +F_{\text{P}}}{P_{\scriptscriptstyle \text{TOF}}+F_{\text{P}}},
\end{equation}
where $\eta = \gamma_{\text{guided}}/\gamma_{\text{total}}$ is the channeling efficiency of the optical nanofiber without mirrors. For an orientation-averaged dipole this means that >80\% of photons are channeled into the cavity. Hence, this resonator is well-suited for increasing the efficiency of single photon sources of solid state emitters with low emission on the zero phonon line \cite{kurtsiefer_photon_source_2000,jungwirth_temperature_hBN_2016,pototschnig_controlling_2011,jeantet_nanotube_2016} and for investigating long-range photon-mediated interactions between different quantum emitters \cite{rist_photon-mediated_2008}.  

\section{Conclusion}

In summary, we present a cryogenic, nanofiber-based microcavity with high quality factors. Together with its small mode volume, this cavity would allow one to significantly increase the coupling between a light field and a quantum emitter by providing a Purcell factor of up to 15. These properties compare very well to those of other micro- \cite{benedikter_cavity-enhanced_2017,faraon_resonant_2011,gallego_strong_2018,schell_cavity_2015,fujiwara_coupling_2012,srinivasan_optical_2007,herzog_cavity_2018,wang_cavity_2019,henze_tuning_2013,Rattenbacher2019} or nanocavities \cite{kaniber_efficient_2007,englund_controlling_2007}. Moreover, our resonator is fully fiber-integrated and alignment-free.
It is therefore suitable for a large variety of emitters \cite{Trusheim2018,gorlitz_tinvacancy_2019,Vamivakas2009,skoff_singlemolecules_2018,Pazzagli2018,Dietrich2018,Serrano2018} and, thanks to its implementation in a cryogenic environment without any loss in transmission, might also be used for the implementation of quantum hybrid systems \cite{minniberger_magnetic_2014,schoelkopf_wiring_2008}.

\section{Appendix A: Temperature-dependent wavelength shift of fiber Bragg grating resonators}

The temperature-dependent wavelength shift of the FBG's reflection band at an initial Bragg wavelength  
$\lambda_{B_0}$ and refractive index $n_0$ is given by \cite{Mizunami2001,Othonos,reid1998,guo_cryogenic_2012}: 
	\begin{equation}\label{eq:wavelengthshift}
	\frac{\mathrm{d}\lambda_B}{\lambda_{B_0}}= (\alpha + \frac{1}{n_0}  \frac{\mathrm{d}n}{\mathrm{d}T} )\ \mathrm{d}T,
	\end{equation}
where $\alpha$ denotes the thermal expansion coefficient and $(1/n_0)  (dn/dT)$ is the thermo-optic coefficient. In our case, two FBGs form a resonator with a center wavelength
$\lambda_\text{c}$, given by the wavelength of best overlap of the reflection bands of the two FBGs. We approximate $\lambda_{\text{c}_0} = (\lambda_{B_{01}}+\lambda_{B_{02}})/2$. Thus, we can apply equation \ref{eq:wavelengthshift} for describing the temperature dependence of $\lambda_\text{c}$.
We measure transmission spectra of the nanofiber-based cavity at room temperature and \SI{4.6}{K}, as shown in Fig. \ref{fig:transmission}. We then  evaluate the shift of $\lambda_\text{c}$. For the nanofiber-based resonator, we find $\lambda_\text{c}= 852.5555 \pm$ \SI{0.0026}{nm} and 851.8944 $\pm$ \SI{0.0025}{nm} for the highest finesse values at room temperature and at \SI{4.6}{K}, respectively. Hence, we measure a shift of -0.6611 $\pm$ \SI{0.0051}{nm} upon cooldown. 

Previously, temperature dependence of FBG's reflection bands has mainly been studied in the context of FBG temperature sensors, where the focus lies on maximising sensitivity by coating the FBG or embedding it into other materials with a high thermal expansion coefficient \cite{habisreuther_FBG_2012, Yamada, reid1998,froevel,Mizunami2001,gupta_bragggrating_1996}. There, the large thermal expansion coefficients of the other materials dominate so that the smaller thermo-optic effect of the FBG itself is less relevant. However, for a silica structure as our nanofiber-based Bragg grating resonator, it is known that the contribution of the thermo-optic effect exceeds that due to thermal expansion \cite{reid1998, white_silica_1975,white_silica_1973,arcizet_microcavities_2009,park_microsphere_2007} when the structure is cooled from room temperature to liquid helium temperatures. At room-temperature, typical values of the thermal expansion coefficient and the thermo-optic coefficient are  $0.55 \times 10^{-6}$ K$^{-1}$ \cite{Othonos} and $6.29 \times 10^{-6}$ K$^{-1}$ \cite{reid1998}, respectively. It is known that the resulting temperature-dependent wavelength shift decreases with decreasing temperature 
\cite{habisreuther_FBG_2012}. We measure a fractional center wavelength-shift of $d\lambda_\text{c}/\lambda_\text{c} = -(7.8\pm1.7) \times 10^{-4}$ for cooling from 295K to 4.6K.  As the effect of $\alpha$ on the total wavelength shift is small compared to the thermo-optic effect, we assume  $\alpha$ to be constant and obtain a fractional change of the refractive index of $dn/n =-(6.1\pm1.7) \times 10^{-4}$. To relate our measurements to previous literature on the thermo-optic coefficient of silica \cite{reid1998,guo_cryogenic_2012,leviton_temperature-dependent_2006,flockhart_quadratic_2004}, one may consider the temperature averaged thermo-optic coefficient of $\overline{\frac{1}{n_0}\frac{dn}{dT}} = \frac{1}{T_f-T_i} \int_{T_i}^{T_f} \frac{1}{n_0}\frac{dn}{dT} dT = (2.1 \pm 0.4) \times 10^{-6}$ K$^{-1}$ for the temperatures $T_i = 295$ K and $T_f = 4.6$ K,  as obtained from our measured shift. The error here is mainly determined by the width of the reflection bands (HWHM $\approx$ \SI{0.1}{nm}). As the overlap of the FBGs may differ between one cavity to the next, the reflection band width determines the precision, with which we are able to measure the thermo-optic coefficient. The effect of using a temperature-dependent value for $\alpha$ instead as has been measured in \cite{white_silica_1973,white_silica_1975} is less than the given error bar.
As a test for our value of the thermo-optic coefficient, we repeat our measurements with a second FBG resonator without a nanofiber-section that has a center wavelength of 852.3305 $\pm$ \SI{0.0097}{nm} at room temperature. We find a shift of -0.7969 $\pm$ \SI{0.0190}{nm}, where the error corresponds to two FSR of the untapered resonator. The corresponding temperature-averaged thermo-optic coefficient is $(2.7 \pm 0.4) \times 10^{-6}$ K$^{-1}$ in good agreement with the first value.

We note that only limited data on thermo-optic coefficients is available for fused silica around \SI{4}{K}. Some studies \cite{leviton_temperature-dependent_2006,flockhart_quadratic_2004} indicate that the temperature-dependence of the thermo-optic coefficient is also wavelength-dependent. This makes more measurements for different wavelengths and especially down to temperatures of \SI{4}{K} valuable for all kinds of future experiments.

\section*{Funding}
Austrian Science Fund (FWF Lise Meitner project No. M 2114-N27);  European
Commission (project ErBeStA No. 800942)

%\section*{Acknowledgments}

%\section*{Disclosures}

%%%%%%%%%%%%%%%%%%%%%%% References %%%%%%%%%%%%%%%%%%%%%%%%%

%%%%%%%%%% If using BibTeX:
\bibliography{paper_citations,quantumBibliography2,cavityBibliography}

\end{document}